\documentclass[reprint,amsmath,amssymb,pra,floatfix,superscriptaddress]{revtex4-1}

\usepackage{bm}
\usepackage{color}
\usepackage[hidelinks]{hyperref}
\usepackage{xcolor}
\hypersetup{
    colorlinks,
    linkcolor={red!50!black},
    citecolor={blue!50!black},
    urlcolor={blue!80!black}
}
\usepackage{graphicx}
\usepackage{epstopdf}
\usepackage{endnotes}

\begin{document}
 \title{Multielectron effects in strong field ionization of CO$_2$: impact on differential photoelectron spectra}
 \author{Vinay Pramod Majety}
 \email[]{vinay-pramod.majety@mpsd.mpg.de}
 \affiliation{Physics Department, Ludwig Maximilians Universit\"at, 80333 Munich, Germany}
 \affiliation{Max Planck Institute for the Structure and Dynamics of Matter, 22761 Hamburg, Germany}
 \author{Armin Scrinzi}
 \email[]{armin.scrinzi@lmu.de}
 \affiliation{Physics Department, Ludwig Maximilians Universit\"at, 80333 Munich, Germany}

\begin{abstract}
  We report fully differential photoelectron spectra from an {\it ab-inito} coupled channels treatment of CO$_2$. Photoionization by laser
  pulses centered at 400 nm and 800 nm wavelength are considered, with arbitrary molecular alignment and polarization (linear and elliptic).
  Calculations reveal significant excited state channel contributions that are, at certain molecular orientations, an order of magnitude larger
  than the ground state channel in the rescattering plateau. Partial wash out of the multiphoton structure in the ATI spectra and of the nodal
  features in angle resolved spectra is observed due to ionization to excited state channels. 
  The qualitative nature of the spectra is determined by ionization thresholds, orbital symmetries and interchannel coupling in the order of precedence.
\end{abstract}
\maketitle

\section{Introduction}

The advent of intense near-IR laser pulses in early 1990's led to the discovery of several highly non-linear phenomena
such as high harmonic generation, high order above threshold ionization and non-sequential multiple ionization \cite{Agostini2012117,scrinzi2006rev,Krausz2009}.
The last two decades saw various proposals of imaging and spectroscopic techniques based on these phenomena that could deepen our
understanding and improve control of atomic-scale phenomena such as chemical reaction mechanisms and solid state transport phenomena.

In the context of molecular imaging, one of the widely pursued techniques is the laser induced electron diffraction (LIED) technique \cite{spanner2004,Lein2007}
where, the electrons ionized by a laser pulse are redirected back to the source by the same laser field where they scatter off the parent
system carrying structural information. Extraction of bond lengths \cite{Meckel13062008,Blaga2012,Pullen2016} and certain dissociation dynamics \cite{Wolter308}
in small molecules from experimentally measured differential photoelectron spectra have been recently reported.
The initial proof of principle experiments show promise, but the evolution of these ideas 
into general techniques would need a deeper understanding of the complex multielectronic and vibrational motion that the strong laser field would induce.

In this paper, we theoretically explore the presence of multielectron effects (within a frozen nuclei approximation), during molecular strong
field ionization and particularly, their impact on photoelectron spectra that would have consequences on LIED imaging.
The challenge to an {\it ab-initio} modeling of molecular strong field ionization (SFI) and computing the associated photoelectron spectra ensues
from the high dimensionality of the problem when several electronic degrees of freedom are allowed to respond to the external field, and from the large phase
space requirements needed to describe strong field ionization, even for a one electron system. Considerable progress has been made to
control the latter problem through the development of the time dependent surface flux method (tSurff) \cite{Tao2012,Scrinzi2012}.
The former challenge however remains. Popular choices to handle this many-body problem in the context of photoionization studies
include the coupled channels method \cite{PhysRevA.80.063411,Majety2015}, the R-matrix method \cite{PhysRevA.79.053411}, variants of the configuration interaction
\cite{PhysRevA.82.023406,Toffoli2016} and the multi-configurational time dependent Hartree-Fock methods \cite{PhysRevA.71.012712,
PhysRevA.86.053424,PhysRevA.89.063416,PhysRevA.91.012509,PhysRevA.91.023417}. Each of these methods cuts down the total Hilbert space to the
seeming essential part needed to describe the ionization process using different protocols.  

The application of these methods to computation of photoelectron spectra in the long wavelength regime has however been
minimal so far. Spectra for atoms, using a multielectron theory, have only been recently reported using the R-matrix method \cite{Hassouneh2015}
and the hybrid anti-symmetrized coupled channels (haCC) method \cite{Majety2015}. In the case of molecules, the increased cost of computations
resulting from reduced symmetry, have rendered such computations infeasible so far. Theoretical modeling has been confined to single electron
approximations and strong field approximation (SFA) based models \cite{Suarez2016,Hetzheim2007,Busulad2008,PhysRevA.95.063410}, with the exception of molecular hydrogen
\cite{Vanne2010,Majety2015}, and the impact of multielectron effects on molecular photoelectron spectra has remained unexplored.

CO$_2$ molecule has often been used in the literature as a prototype for non-polar molecules. It is one of the simplest molecules that exhibits signatures of multielectron
effects in various observables associated with strong field ionization (SFI). In a seminal work by Smirnova and co workers \cite{Smirnova2009}, the relevance of
inner orbital ionization and channel coupling in high harmonic spectroscopy of CO$_2$ was demonstrated. In \cite{dyn_exch}, again using CO$_2$ as an example
, the role of exchange interaction in SFI was demonstrated by the authors.
We report here first fully differential photoelectron spectra for CO$_2$ molecule in the low frequency regime, obtained from a
multielectron approach and analyze the nature of multielectron effects involved. We use the hybrid anti-symmetrized coupled channels treatment of
the molecular electronic structure presented in \cite{Majety2015} and established in \cite{dyn_exch,sfir} with the tSurff to compute fully
differential (angle and channel resolved) photoelectron spectra. Spectra are computed with linear as well as elliptically polarized ionizing fields.

The fully differential photoelectron spectra presented here reveal significant excited ionization channel contributions
in consonance with the earlier findings on the importance of inner orbital ionization \cite{Smirnova2009}.
In addition, these excited state channels dominate in the rescattering region depending on the molecular orientation with respect 
to the laser polarization (defined by an angle, say $\alpha$). For example, when $\alpha=90^\circ$, the highest occupied molecular orbital (HOMO)-1
channel dominates HOMO channel by an order of magnitude in the rescattering region. This opens up the possibility to image channel specific dynamics in LIED
experiments by choosing a specific molecular orientation.
Other consequences of multiple orbital ionization include wash-out of the multiphoton peak structure in certain photoelectron electron energy windows and of
the nodal features in the total angle resolved photoelectron spectra. An ionic channel analysis of the spectra and its comparison with the corresponding
single channel calculations reveal that the spectra are qualitatively determined by the ionization thresholds, orbital symmetries and interchannel couplings
in the order of precedence.

We present the computational methodology in brief in section \ref{sec:method} and discuss the results in section \ref{sec:results}.

\section{Computational method} \label{sec:method}

The molecular time dependent Schr\"odinger equation is solved within the frozen nuclei approximation with the following
basis expansion for the electronic wavefunction:
\begin{equation} \label{method:basis}
 | \Psi \rangle = \sum_{i,\mathcal{I}} \mathcal{A} |i\rangle | \mathcal{I}\rangle C_{i,\mathcal{I}}(t) +  |\mathcal{G} \rangle C_{\mathcal{G}}(t).
\end{equation}
Here, $\mathcal{A}$ indicates anti-symmetrization, $| \mathcal{I}\rangle$ are a set of singly charged cationic states, $|\mathcal{G} \rangle$ is the ground state of 
the neutral, $|i\rangle$ is a complete single electron basis and $C_{i,\mathcal{I}}(t)$, $C_{\mathcal{G}}(t)$ are time dependent coefficients corresponding
to their respective basis functions. A calculation with lowest I-channels is referred to as ``haCC(I)'' calculation and degenerate states are counted separately.
A summary of the method is provided here, while the detailed description can be found in \cite{Majety2015}. 

The single electron basis $|i\rangle$ is a product basis consisting of high order polynomials on finite elements for the radial coordinates and spherical
harmonics based on the origin for the angular coordinates. The ionic and the neutral states are computed using multi-reference configuration interaction
singles doubles method implemented in the COLUMBUS quantum chemistry package \cite{Lischka2011} using atom centered Gaussian basis set (minimally augmented cc-pvtz basis)
at the starting point. The choice implies that the multi-centered nature of the initial state and the residual ionic states is represented accurately with the Gaussians, leaving
only the single continuum to be described by $|i\rangle$. As a result, the angular momentum requirements in the single center expansion $|i\rangle$ remain
at par with a corresponding atomic case.
On a 25 a.u. simulation volume, we obtain angle integrated photoelectron spectra that have a relative accuracy of about 10\% with respect to radial basis
parameters and angular momenta with 50 radial functions constructed from order 10 polynomials and with 35 angular momenta.

\begin{figure}[htbp]
\begin{minipage}{0.24\linewidth}
   \centering
   \includegraphics[width=0.65\linewidth]{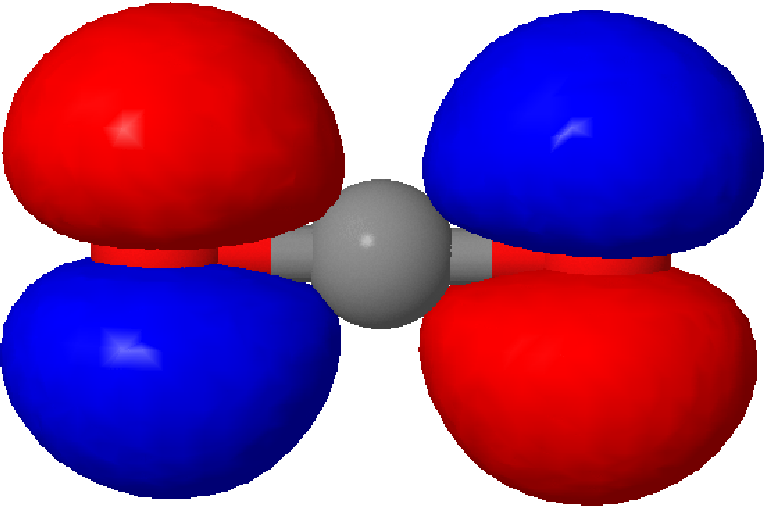}       
\end{minipage}
\begin{minipage}{0.24\linewidth}
   \centering
   \includegraphics[width=0.65\linewidth]{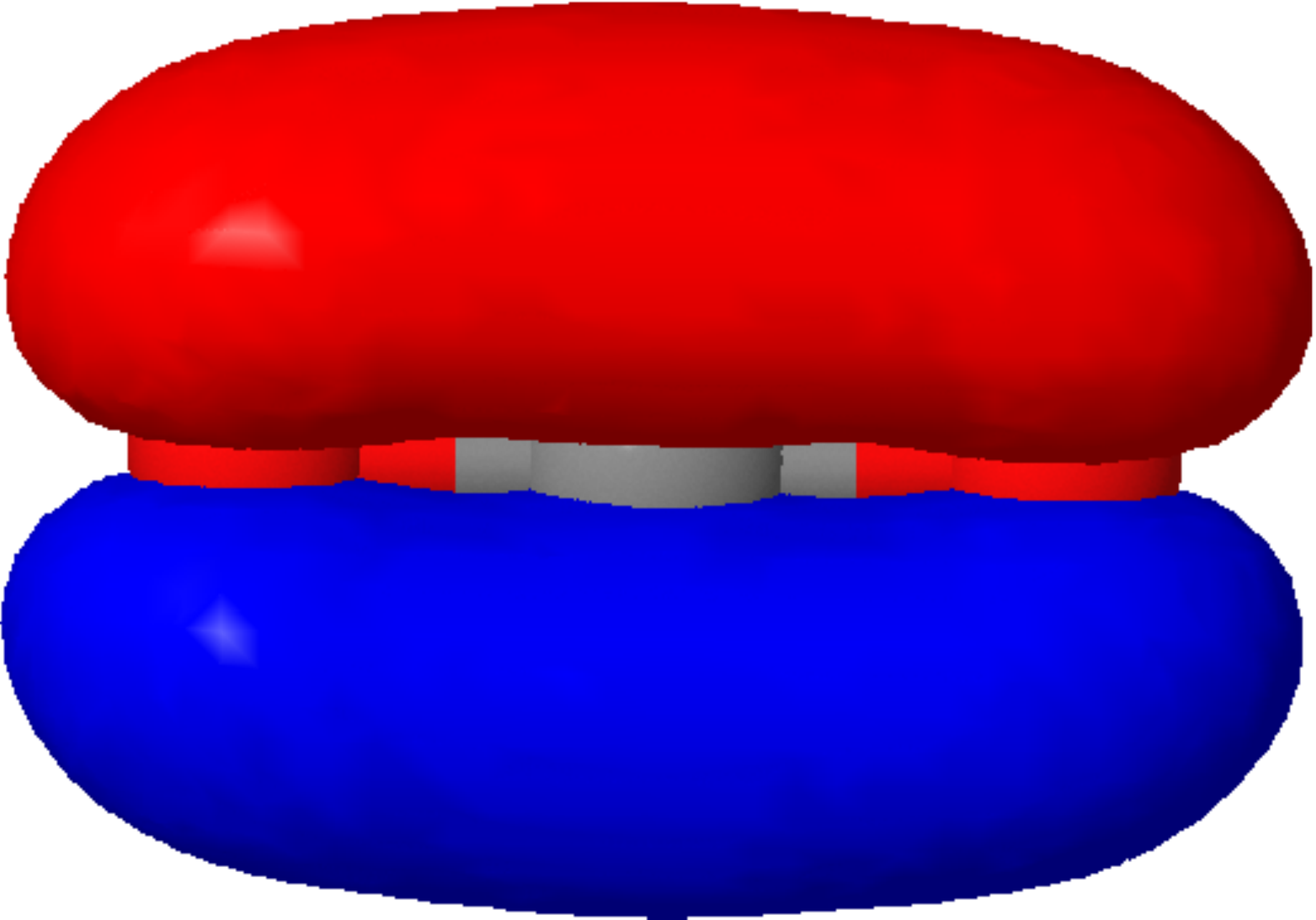}         
\end{minipage}
\begin{minipage}{0.24\linewidth}
   \centering
   \includegraphics[width=0.75\linewidth]{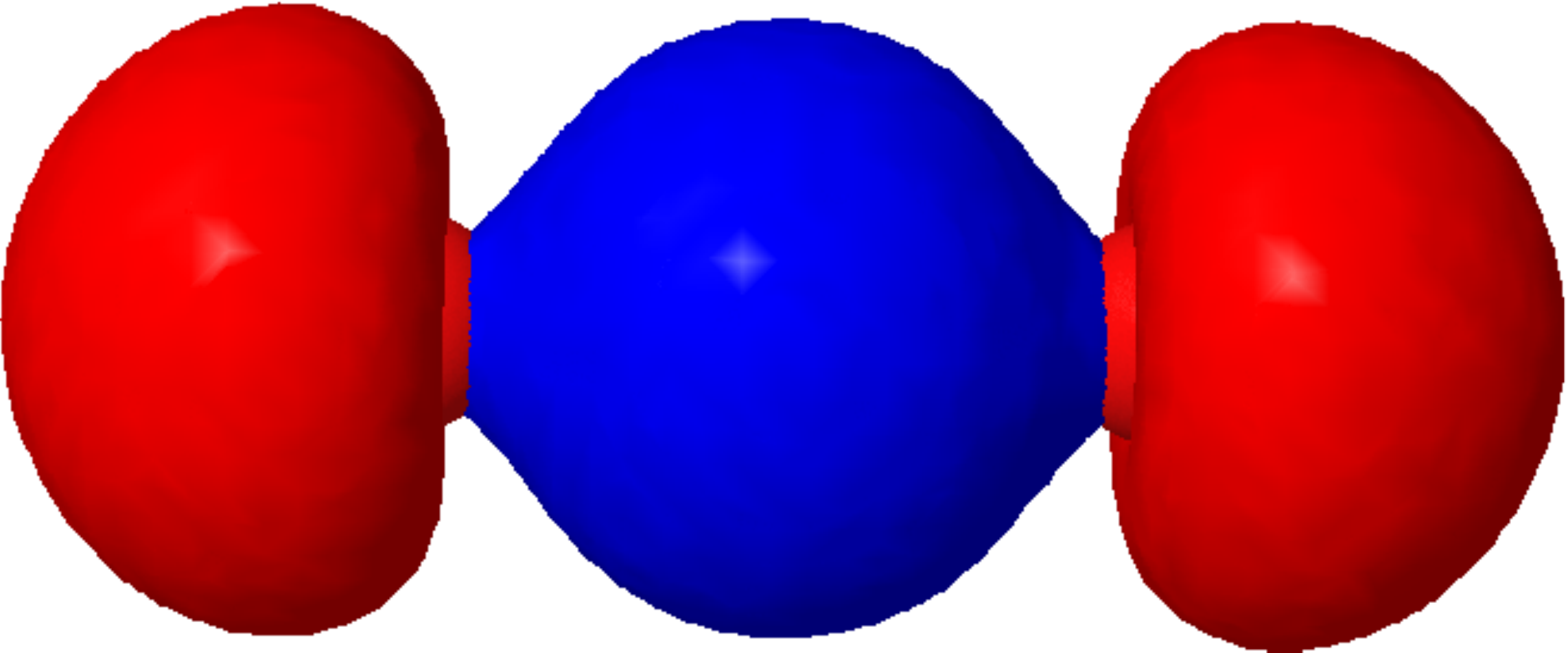}       
\end{minipage}
\begin{minipage}{0.245\linewidth}
   \centering
   \includegraphics[width=0.75\linewidth]{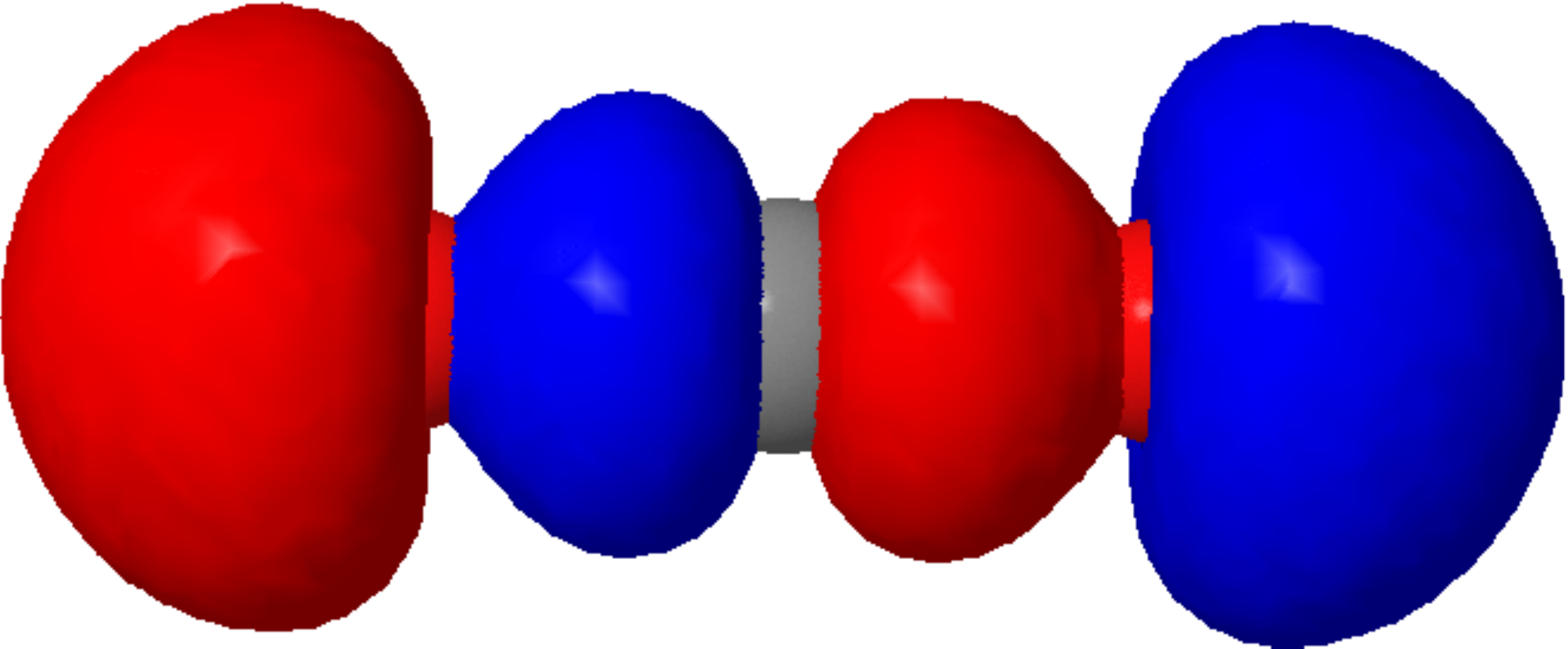}         
\end{minipage}

\caption{Isosurfaces of the highest occupied molecular orbital (HOMO), HOMO-1, HOMO-2 and HOMO-3 ordered from left to right, plotted with Jmol software
 \cite{jmol} using the Hartree Fock output of our COLUMBUS calculations.}
\label{fig:orbitals} 
\end{figure}

We considered CO$_2$ molecule frozen at equilibrium configuration with C-O bond length of 116.3pm and the single cationic states $X^2\Pi_g$, $A^2\Pi_u$, $B^2\Sigma_u^+$,
$C^2\Sigma_g^+$ and $D^2\Pi_u$ states with vertical binding energies of 13.68 (13.78, 13.77), 17.74 (17.59, 17.72), 18.46 (18.08, 18.28),
19.74 (19.4, 19.48) and 25.47 (22.7, 24.19) eV respectively with respect to $|\mathcal{G}\rangle$. Values in brackets are the experimental (first) and theoretical (second), as reported in \cite{Ehara1999}.
The first few highest occupied molecular orbitals on which the CI states are based on are shown in figure \ref{fig:orbitals} as a reference to orbital symmetries 
needed to analyze our results.

When diagonalizing with the coupled channels basis, the ground state improves and the thresholds increase. The first ionization potential varies from 13.7 eV to 13.88 eV
between the smallest to the largest coupled channels basis considered here. 
The change in the ionization potential (IP) between various haCC calculations is $\leq$12\% of the photon energies considered. Hence, the possibility for the order of 
the multiphoton ionization process to change due to the shift in the IP is minimal. Ionization by tunneling is highly sensitive to the IP.
An estimate using MO-ADK theory \cite{moadk} suitable under the quasistatic limit shows that the ionization rates would change on the scale of 20\% for a change in the
IP by 0.18 eV. This is within the relative accuracy for the spectra that we aim here. Other examples where IP is crucial would be processes
involving resonances. We do not deal with such processes here.

The ionizing field is approximated to have a $\cos^2$, $\cos^8$ envelopes for linearly and elliptically ($\epsilon=0.87$) polarized fields, respectively.
In both cases, we consider 3-cycle pulses with central wavelengths of 400 and 800 nms. The peak intensities were chosen to be 10$^{14}$ W/cm$^2$ in the linear
case and 1.76$\times$10$^{14}$ W/cm$^2$ in the elliptic case. This sets the peak field strengths to the same value in both cases.
The molecular axis is fixed to be along z axis. With linear polarization, polarization directions parallel ($\alpha=0^\circ$ ), perpendicular ($\alpha=90^\circ$) and at $45^\circ$
with respect to the molecular axis are considered. In the elliptic case, fields in the plane parallel and perpendicular to
the molecular axis are considered. The exact forms for the vector potentials are:
\begin{equation}
 A(t) = A_0 \cos^2(\frac{\pi t}{2N_cT}) \sin(\omega t) \hat{p}
\end{equation}
and
\begin{equation}
 A(t) = \frac{A_0}{\sqrt{1+\epsilon^2}} \cos^8(\frac{\pi t}{4N_cT}) \left( \sin(\frac{2\pi t}{T}) \hat{e}_1 + \epsilon\cos(\frac{2\pi t}{T}) \hat{e}_2 \right)
\end{equation}
where $A_0$ is the vector potential at the peak field, $T$ is the duration of the optical cycle at the respective wavelength, $\epsilon$ is the measure of the ellipticity,
$N_c$ is the number of cycles and $\hat{p},\hat{e}_1,\hat{e}_2$ are unit vectors that define the polarization. When the field is in the XZ plane,
$(\hat{e}_1,\hat{e}_2) = (\hat{z},\hat{x})$, and when in the XY plane, $(\hat{e}_1,\hat{e}_2) = (\hat{y},\hat{x})$

The non-perturbative external laser field is treated classically in the dipole approximation and the wavefunction is discretized in a mixed gauge
representation. Length gauge representation is used in the region encompassing the ionic, neutral states and a velocity gauge
representation thereafter. The specific gauge choice is motivated by our earlier findings on its efficiency \cite{mixedg}.
The time propagation is done with the classical fourth order Runge-Kutta method and the absorbing boundary conditions are
imposed using the infinite range exterior complex scaling scheme \cite{PhysRevA.81.053845}.

Fully differential photoelectron spectra are computed using the time dependent surface flux method (tSurff) \cite{Tao2012,Scrinzi2012}, where
the spectra are computed from the electron flux through a surface (defined by a radius referred to here as the tSurff radius ($R_c$)),
instead of a brute force projection of the wavefunction over the entire volume onto the single continuum that are, anyway, only approximately known.
As a result, the wavefunction is represented numerically only on a small spherical region where non-trivial dynamics (other than the motion of a free electron) occur and
absorbing boundary conditions are imposed at $R_c$. Beyond that radius, the dynamics is assumed to follow the Volkov solutions\cite{Alej2016} for a free
electron in a dipole field. This amounts to truncating all the potentials to zero beyond $R_c$ and to avoid artifacts caused
by the abrupt truncation of the potential, we smoothly truncate the potential using a third order polynomial over a small region before $R_c$.
The approximation can be controlled by moving the tSurff radius outwards and an error estimate can be made.

The truncation affects the Hartree potential and nuclear Coulomb potential whose combination far from nucleus has a $1/r$ behavior. The exchange potentials,
as they involve overlaps with the Gaussian bound orbitals, fall off even more quickly and are not affected by this truncation. (For matrix elements involved, see \cite{Majety2015}).

In this work, we do not aim for a rigorous convergence with respect to the long range Coulomb truncation. In order to keep the computational
requirements under check, we restrict ourselves to a simulation volume of radius 25 a.u that captures the essential multi-electron dynamics.

The largest computations presented here took about 12 days on a standard 16 core machine with 128 GB of memory using 
our current implementation for shared memory machines.

\subsection{Limitations}
The ansatz (\ref{method:basis}) treats polarization contribution from the single continuum and multi-electron effects such as
exchange interaction and channel couplings exactly. But the polarization of the core electrons is not described completely. In principle,
a complete description of polarization for all the electrons would need the complete continuum space and not just the single continuum. This, however,
entails to solving the Schr\"odinger equation in full generality which is not feasible. Conceptually, one can extend the current formulation to
include double continuum states, but this could not been attempted so far due to increased computational costs.

A coupled channels model is a useful tool to study molecular strong field ionization only when the relevant multi-electron effects can be described
using a few channel expansion. One would expect that the current regime of ionization, that is highly nonlinear with respect to ionization thresholds,
would fit into such a case if the core electrons are only ``gently" polarized. For obtaining ionization rates, such a description was found to be sufficient
\cite{dyn_exch,sfir}, though it remains to be seen if it holds in the case of a differential quantity such as photoelectron spectra.

If a large number of highly excited states are required, our model ceases to be suitable. Increasing the number of states beyond the lowest few does
not necessarily lead to convergence, as the ionic bound states do not form a complete set. Further, the highly excited states would need to be strongly
adjusted to the field in order to maintain any specific physical meaning over an arbitrarily chosen basis. Hence, convergence in a strict sense
using a large number of ionic channels is not sought for, here. The model can however allow us to understand the role of possible multi-electron effects by
studying the dependence of spectra on the few channels that can be included in practice. In this work, we restrict ourselves to the first 6-8 ionic
channels.

\begin{figure*}[htbp]
\begin{minipage}{0.32\linewidth}
   \centering
   \includegraphics[width=0.98\linewidth]{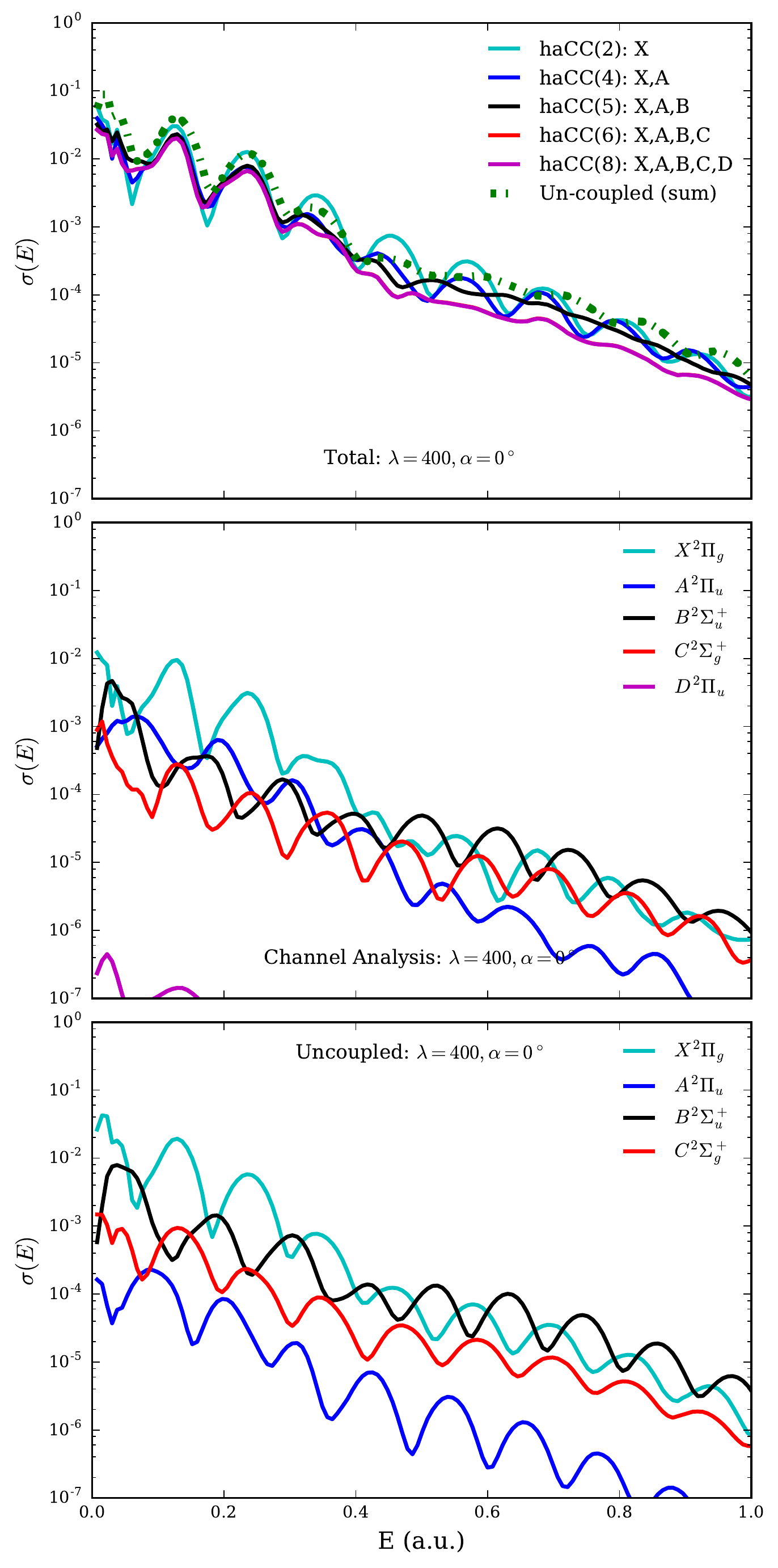}          
\end{minipage}
\begin{minipage}{0.32\linewidth}
   \centering
   \includegraphics[width=0.98\linewidth]{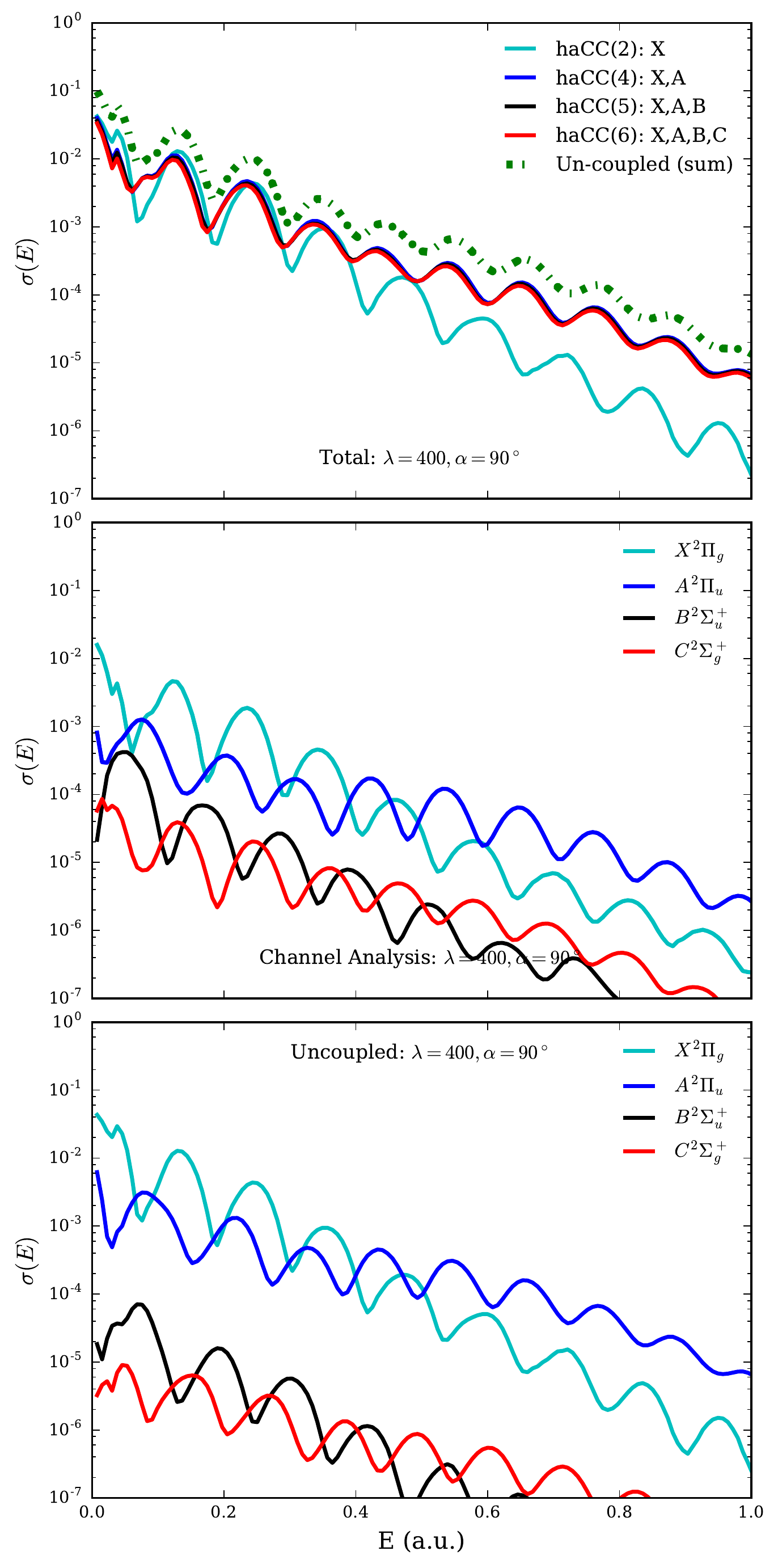}         
\end{minipage}
\begin{minipage}{0.32\linewidth}
   \centering
   \includegraphics[width=0.98\linewidth]{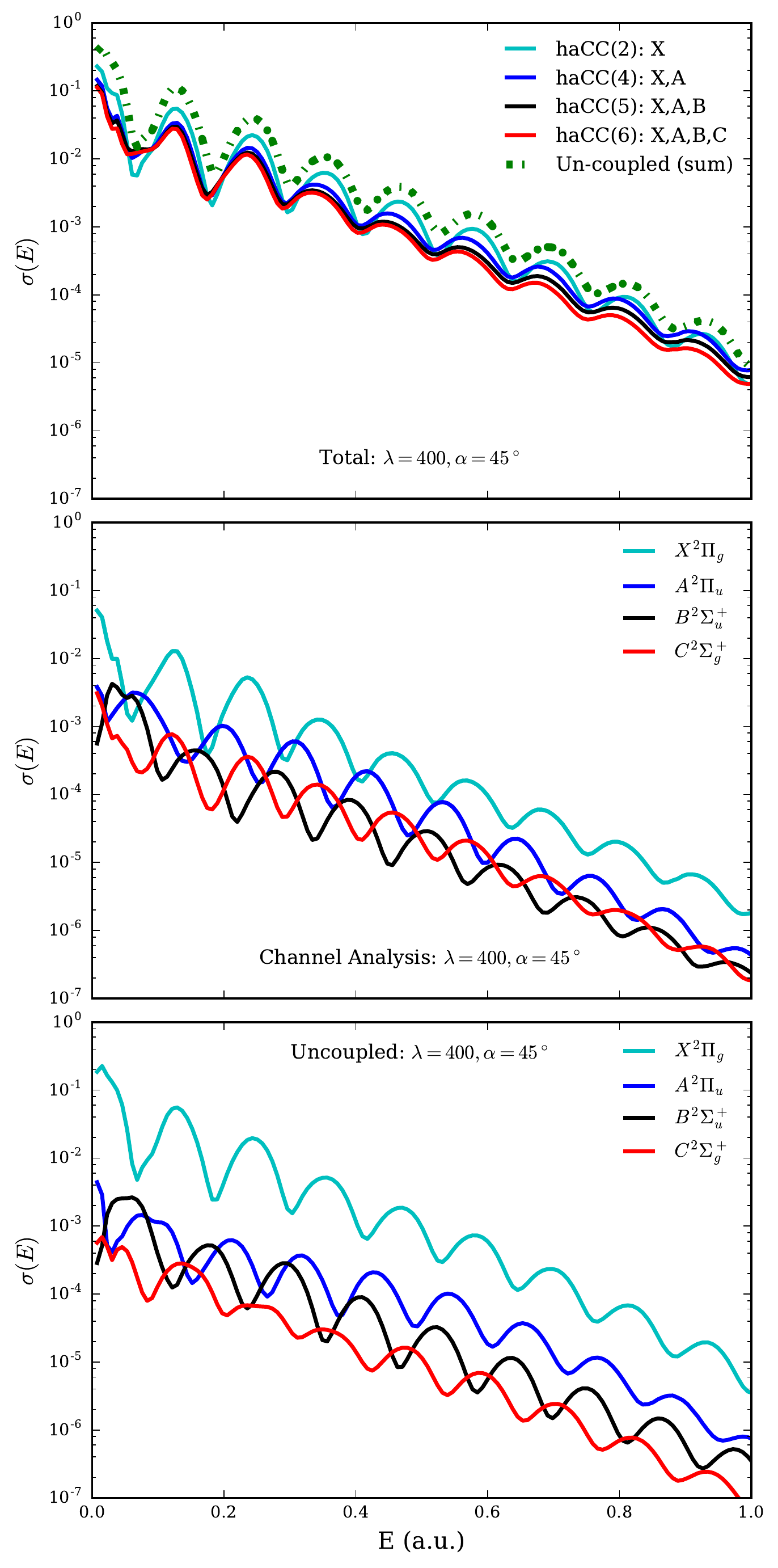}         
\end{minipage}
\caption{Single ionization spectra at 400 nm with molecular orientations $\alpha=0^\circ$ (left column), $\alpha=90^\circ$ (center column) and $\alpha=45^\circ$ (right column).
Plots in the upper row show spectra from coupled channels calculations with different number of ionic channels and sum spectra from uncoupled channel calculations.
The plots in the second row show ionic channel analysis of the spectra from the largest haCC calculation and the plots in the bottom panel show spectra from uncoupled single
channel calculations.}
\label{fig:400}
\end{figure*}

\subsection{Uncoupled calculations}

In order to analyze the nature of multielectron effects that appear in the spectra, we also perform uncoupled single channel calculations
where the wavefunction is represented as
\begin{equation} \label{method:basis_uncpld}
 | \Phi_\mathcal{I} \rangle = \sum_{i} \mathcal{A} |i\rangle | \mathcal{I}\rangle C_{i,\mathcal{I}}(t) +  |\mathcal{G} \rangle C_{\mathcal{G}}(t).
\end{equation}
These uncoupled calculations correspond to single electron calculations when only the Dyson orbital
($\langle \mathcal{I} | \mathcal{G} \rangle$) corresponding to the considered channel $|\mathcal{I}\rangle$ is allowed to ionize. Note, exchange however is
fully respected, which is different from single electron calculations in model potentials.

These single channel functions, $\Phi_\mathcal{I}$, are only orthogonal asymptotically when one electron is far from the nucleus. Close to the nuclei, they
are non-orthogonal. In addition, each calculation contains the neutral ground state which introduces further non-orthogonality. As a consequence, we cannot
rigorously consider this as a decomposition into non-interacting channels. Quantitative comparisons, specially for the sum spectra, between uncoupled channel
calculations and full haCC calculations must be treated with caution, due to possible double counting.

\subsection{Observables}

Here, we give the explicit forms of the observables presented in the paper. We label the four fold differential photoelectron spectrum as $\sigma(\mathcal{I},E,\Omega)$,
where $\mathcal{I}$ is the residual ionic channel index, $E$ is the energy of the photoelectron and $\Omega=(\theta,\phi)$ denotes the angles defining the direction of emission. The
total integral of this object gives ionization yield, $Y$
\begin{equation}
 Y = \sum_{\mathcal{I}} Y_\mathcal{I} = \sum_{\mathcal{I}} \int \int dE\;d\Omega\;\sigma(\mathcal{I},E,\Omega)
\end{equation}
with $Y_\mathcal{I}$ being the ionization yield into each residual ionic channel.
The energy resolved spectrum is defined as
\begin{equation} \label{eq:specamp}
\sigma(E) = \sum_{\mathcal{I}} \int d\Omega\;\sigma(\mathcal{I},E,\Omega).
\end{equation}
and the energy, channel resolved spectrum is defined as
\begin{equation}
\sigma_\mathcal{J}(E) = \int d\Omega\;\sigma(\mathcal{J},E,\Omega).
\end{equation}

With arbitrary polarizations, symmetries are lost and hence we present various cuts (in the 3 dimensional space) of the angle, energy resolved spectrum.
Such a cut, for example, in the XY plane can be defined as:
\begin{equation}
 \sigma(E,\theta=\frac{\pi}{2},\phi \in [0:2\pi) ) = \sum_\mathcal{I} \sigma(\mathcal{I},E,\Omega)
\end{equation}

\begin{figure*}[htbp]
\begin{minipage}{0.32\linewidth}
   \centering
   \includegraphics[width=0.98\linewidth]{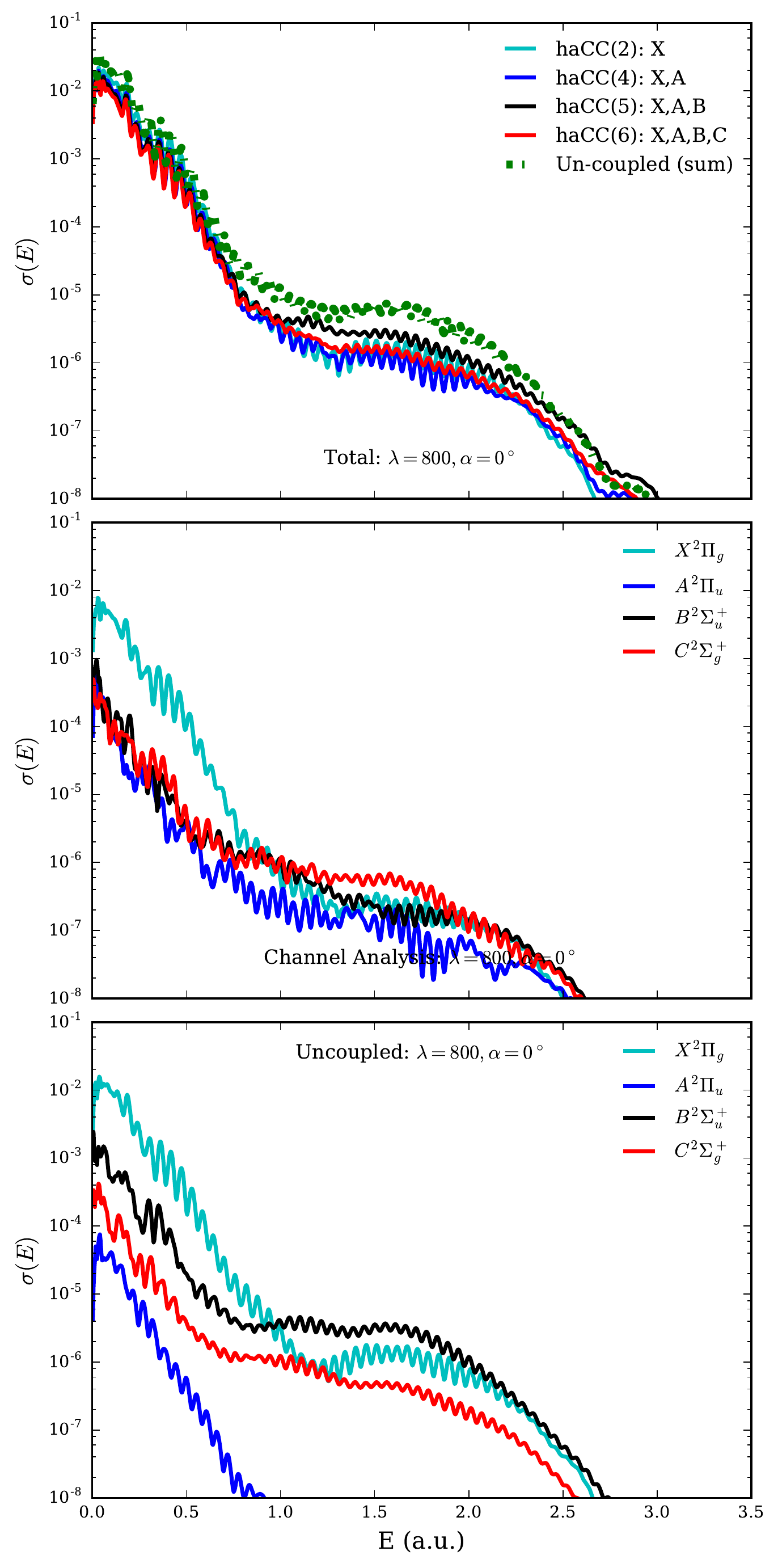}          
\end{minipage}
\begin{minipage}{0.32\linewidth}
   \centering
   \includegraphics[width=0.98\linewidth]{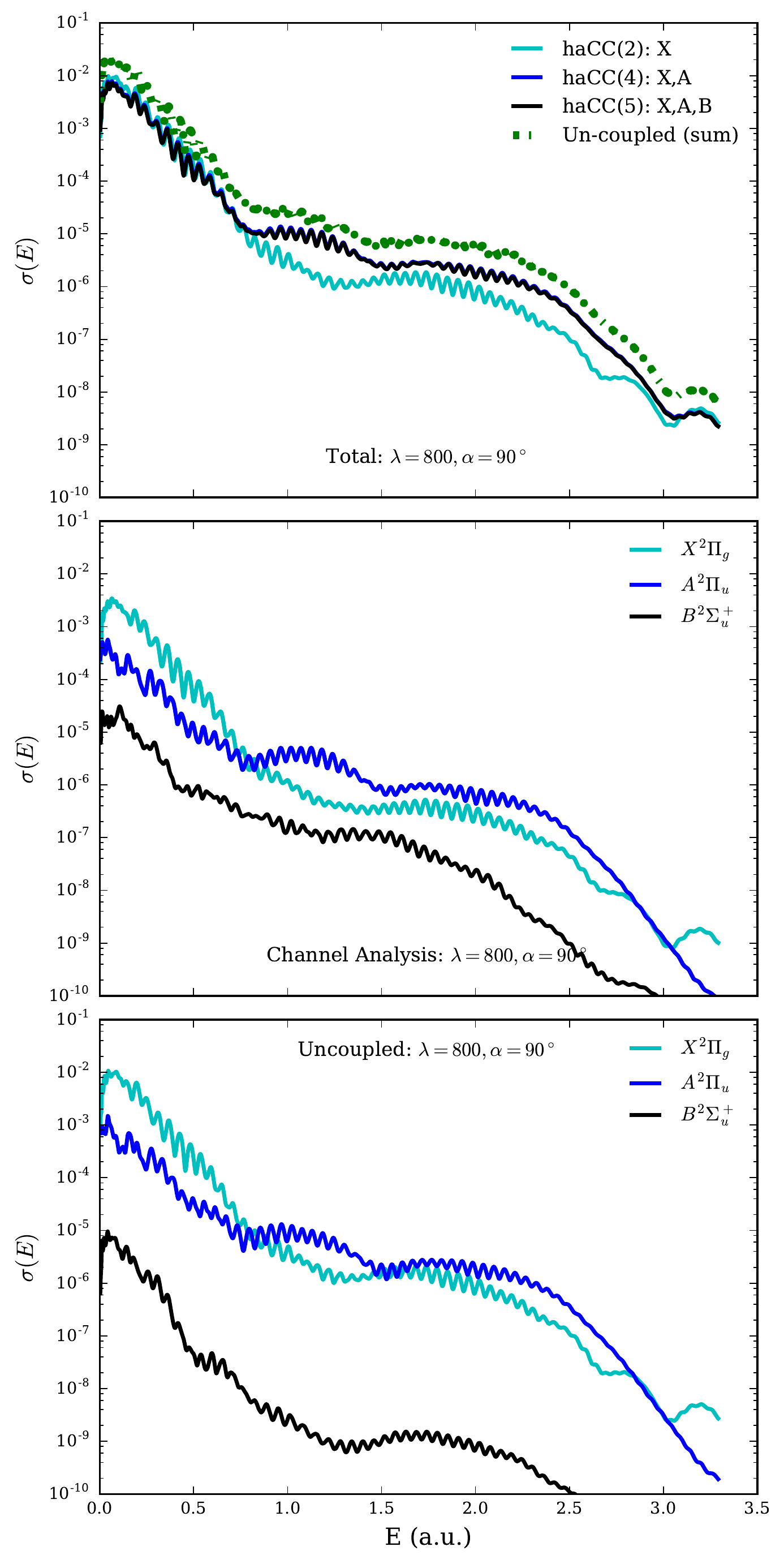}         
\end{minipage}
\begin{minipage}{0.32\linewidth}
   \centering
   \includegraphics[width=0.98\linewidth]{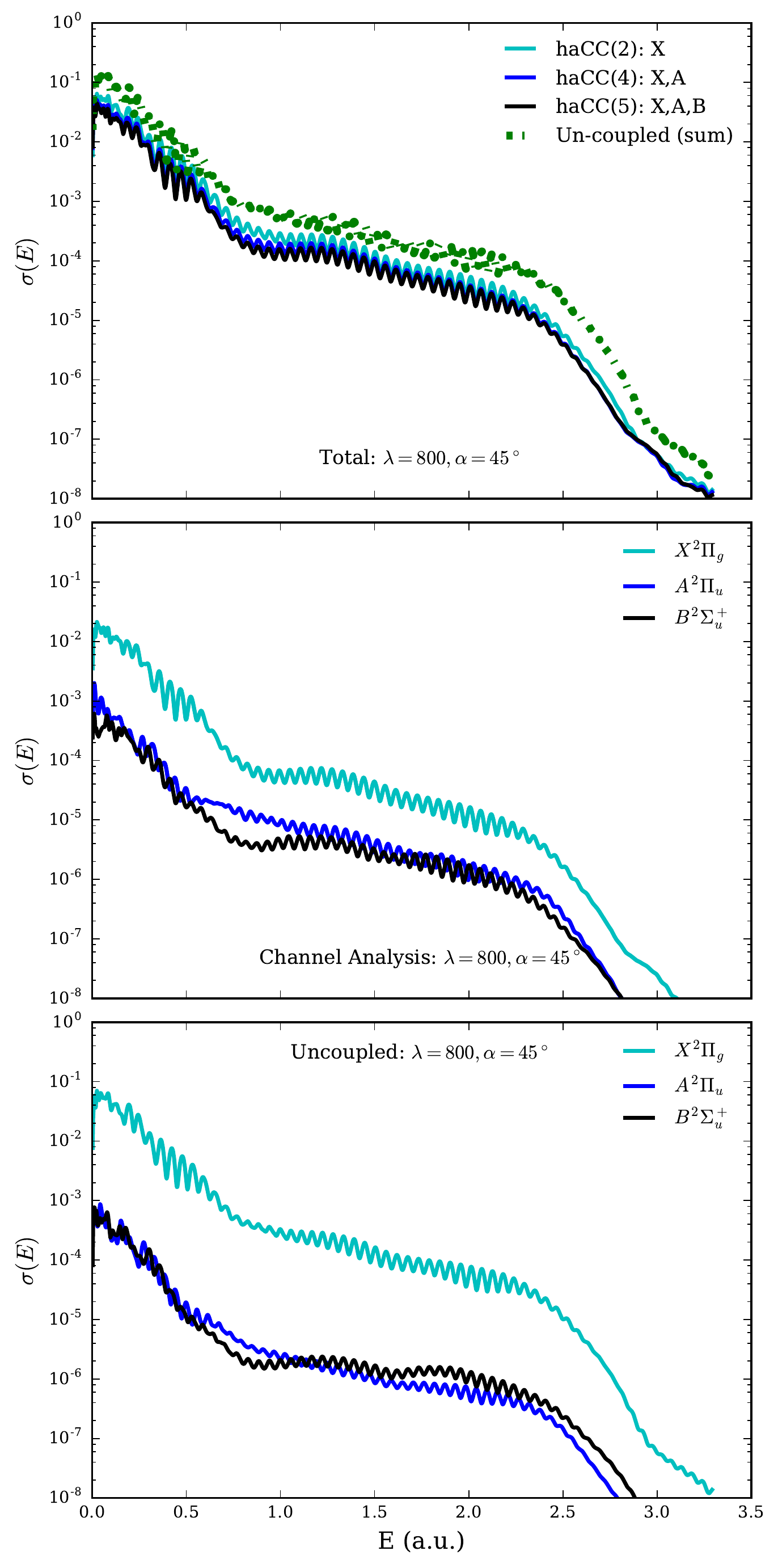}         
\end{minipage}
\caption{Single ionization spectra at 800 nm with molecular orientations $\alpha=0^\circ$ (left column), $\alpha=90^\circ$ (center column) and $\alpha=45^\circ$ (right column).
Plots in the upper row show spectra from coupled channels calculations with different number of ionic channels and sum spectra from uncoupled channel calculations.
The plots in the second row show ionic channel analysis of the spectra from the largest haCC calculation and the plots in the bottom panel show spectra from uncoupled single
channel calculations.}
\label{fig:800}
\end{figure*}

\section{Results} \label{sec:results}

In this section, we present photoelectron spectra from coupled channels calculations and analyze the nature and relevance of multielectron effects
by performing ionic channel analysis and by comparing with uncoupled single channel calculations.

\subsection{Linearly polarized ionizing fields} 

The upper panels of figures \ref{fig:400}, \ref{fig:800} show total single ionization spectra as a function of number of ionic channels in the coupled
channels expansion at 400 nm, 800 nm respectively and with polarization directions defined by angles $\alpha = 0^\circ, 90^\circ$ and $45^\circ$. 

The largest calculations that were possible with the current implementation are presented in each case. With laser parameters [$\lambda=$400 nm, $\alpha=0^\circ$],
upto 8 channels could be included, with [$\lambda=$800 nm, $\alpha=0^\circ$] and [$\lambda=$400 nm, $\alpha=90^\circ, 45^\circ$] up to 6 channels and for 
the cases of [$\lambda=$800 nm, $\alpha=45^\circ, 90^\circ$] up to 5 channels could be included.
The largest calculations that could be performed in each case was defined by the duration of the ionizing pulse and the symmetry of the problem.
With the exception of the spectra at parameters [$\lambda=$800 nm, $\alpha=0^\circ$], the spectra amplitudes (Eq \ref{eq:specamp}) from largest 2 coupled channels
calculations in each case vary by less than 30\%. Based on the impact of the $D$ channel at [$\lambda=$400 nm, $\alpha=0^\circ$], one does not expect further changes
in [$\lambda=$800 nm, $\alpha=0^\circ$] spectrum.

The spectra exhibit above threshold ionization (ATI) peaks. At 400 nm, the ATI peaks fall off exponentially with the order. At 800 nm, the spectra exhibit
the well known rescattering plateau with characteristic cut-offs around $2U_p$ and $10 U_p$ ($U_p \approx 0.22$ a.u), where $U_p$ is the pondermotive energy.
In all of the cases, the spectra change with the inclusion of excited ionic channels indicating the participation of more than one
electron in the ionization process. The total yields decrease when the number of channels is increased. 
In addition to the improvement in the field free ground state, the dynamical core electron polarization is better described by the larger channel basis which leads to
larger quadratic stark shifts of neutral versus the residual ion. Hence, along with the 0.18 eV increase in the field free ionization potential from haCC(2)
to haCC(6), there is an additional increase in the Stark shift of the ground state by about 0.2 eV.
The net increase in the ionization potentials leads to a general reduction in ionization
probability as is expected in the current regime of tunnel and multiphoton ionization
(in the absence of any resonances). This is consistent with the findings in \cite{dyn_exch}, where we presented static field ionization rates. On a
qualitative level, one observes the following as the channel basis is enlarged:  The changes are minimal when  $\alpha=45^\circ$. When $\alpha=0^\circ$,
a wash out of the multiphoton peak structure is observed in the photoelectron energy range 0.4-1.0 a.u at 400 nm and in the energy range 1-1.5 a.u at 800 nm.
When $\alpha = 90^\circ$, there is a significant enhancement of the yields at higher photoelectron energies.

To better understand the spectral features, we perform an ionic channel analysis and preform uncoupled channel calculations according to ansatz (\ref{method:basis_uncpld}).
The corresponding results for each case are presented in the second and third rows of the figures \ref{fig:400} and \ref{fig:800}.
The sum of the uncoupled channel calculations is presented in the top panels. Qualitatively, the sum of the uncoupled spectra agrees with the full haCC spectrum,
but consistently exceeds it. This might be attributed to the absence of any multi-electron Stark shift. However, a rigorous discussion would require ruling out
any double counting due to the non-orthogonality of the uncoupled channels.

Ionic channel analysis reveals that the contribution of excited channels to the total ionization yield in all considered cases is on the scale of 5-10\%
indicating that they must be considered for any careful analysis. While at $\alpha=45^\circ$, the ground state channel dominates at all photoelectron energies,
at $\alpha=0^\circ$, $B,C$ states and at $\alpha=90^\circ$, $A$ state dominate at higher photoelectron energies. The wash out of the multiphoton peak structure
occurs as several ionic channels contribute on a similar level at the concerned photoelectron energies and the ionization thresholds that vary by only a few eV
cause different offsets of the multiphoton peaks in the various channels.

\begin{figure*}[htbp]
\begin{minipage}{0.32\linewidth}
   \centering
   \includegraphics[width=0.98\linewidth]{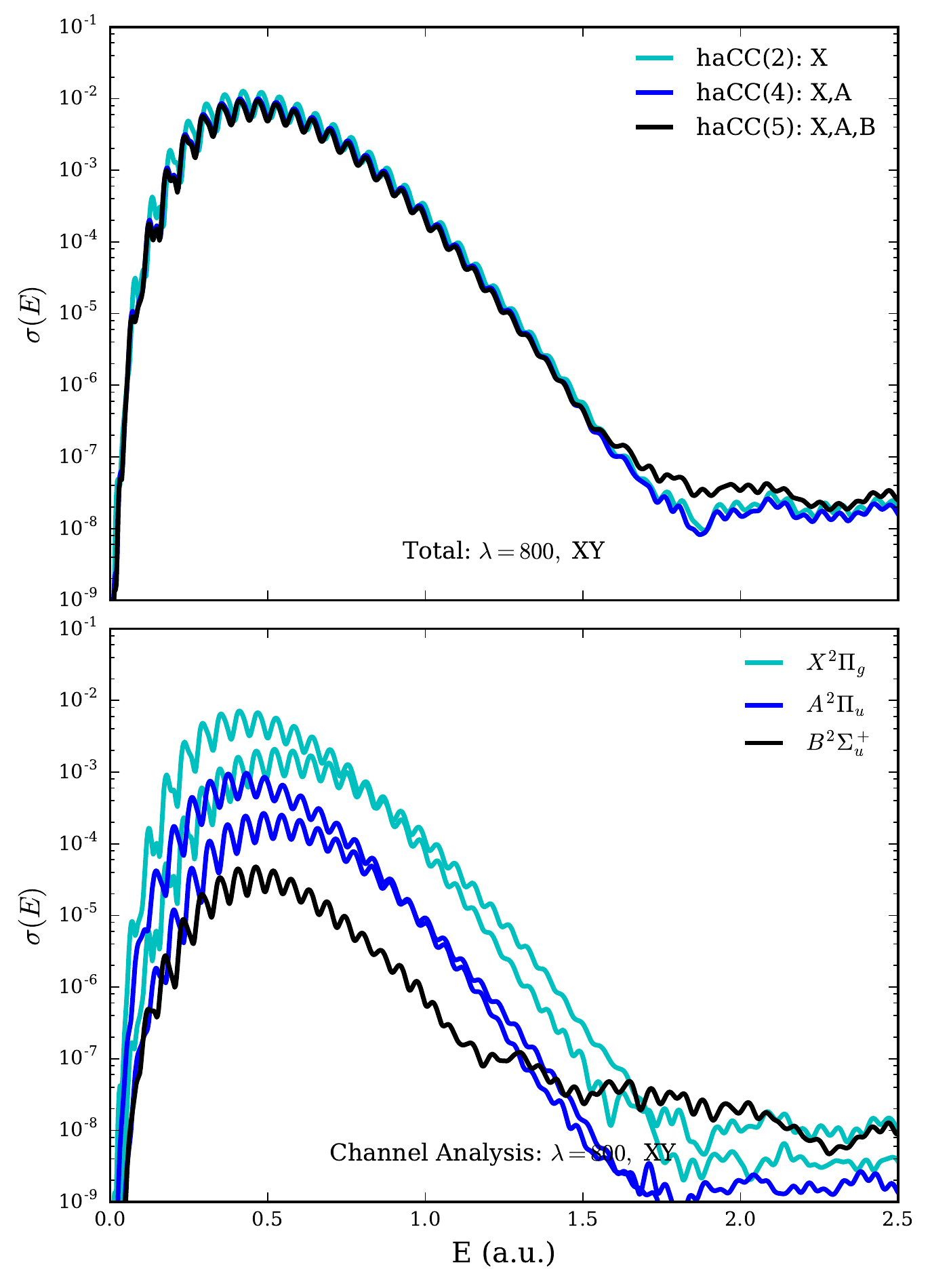}          
\end{minipage}
\begin{minipage}{0.32\linewidth}
   \centering
   \includegraphics[width=0.98\linewidth]{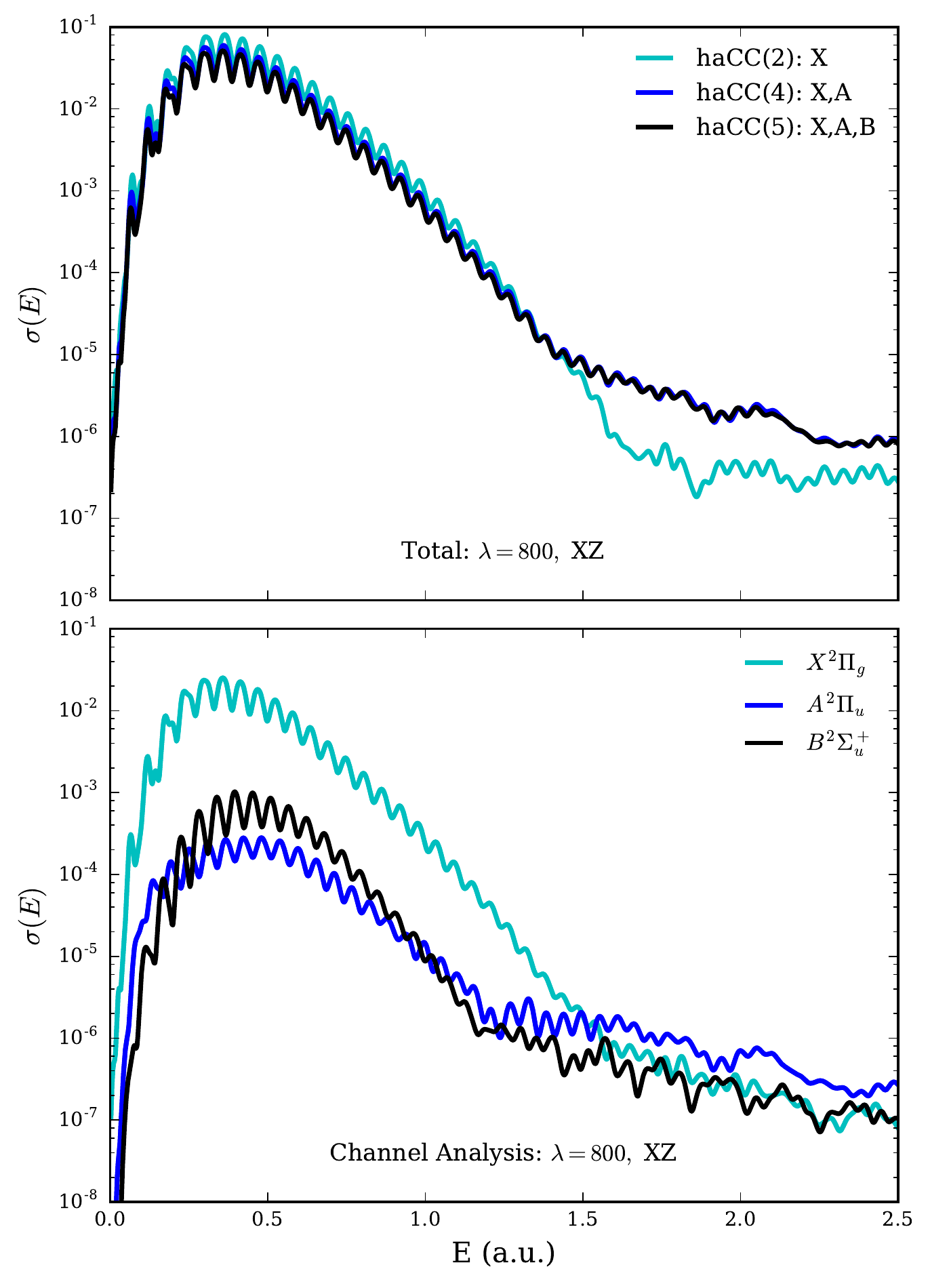}         
\end{minipage}
\caption{Single ionization spectra at 800 nm with elliptic polarization in XY (left) and XZ (right) planes. Plots in the upper row show spectra from
coupled channels calculations with different number of ionic channels. The plots in the lower row show ionic channel analysis of the spectra from the
largest haCC calculation. When ionizing field is in XY plane, the symmetry of the $\Pi$ states is broken and the ionization contributions differ indicated by curves in same color.}
\label{fig:800ellip}
\end{figure*}

This unexpected dominance of the excited ionization channels at higher photoelectron energies is a result of the
symmetry of molecular orbitals: interplay of the nodal planes and the electron density. At 800 nm, the 3-step classical model \cite{PhysRevLett.71.1994} can be employed to
understand the enhancement of excited states in the plateau region which has been well established as a signature of rescattering.
This is further supported by the absence of a plateau structure with elliptically polarized fields presented in the next section.
When $\alpha=90^\circ$, the $A$ channel dominates as the ground state $X$ channel has a node in the polarization plane in which rescattering occurs, where as,
when $\alpha=0^\circ$ degrees, both $X$ and $A$ channels are suppressed and ionization to $B$, $C$ channels become important.
Since the ionization threshold to different channels only differ by few eV, suppression of ionization to a specific channel for symmetry reasons
uncovers contributions to ionization to excited state channels.
At 400 nm, the separation of ionization into tunneling and recollision looses validity and these pictures cannot be applied uncritically.
The similarity of the slopes of the channel resolved spectra and the uncoupled single channel
spectra, indicates that the behavior can be attributed to the nature of the ionizing orbital that determines the multiphoton ionization cross-sections and
thereby the slope of the curves.

The uncoupled single channel calculations, qualitatively, are similar to the channel resolved spectra at $\alpha=45^\circ$
and $\alpha=90^\circ$. This indicates that ionization at these parameters can qualitatively be considered as largely independent ionization into the
individual channels.
At $\alpha=0^\circ$, however, the relative contributions of $B$, $C$ channels in the coupled channels calculation versus uncoupled
single channel calculations differ. This effect is more striking at 800 nm compared to 400 nm case.
Both states share the same nodal structure and are energetically separated by about 1 eV. In the uncoupled calculation, the lower lying $C$ state
contributes less, while in fully coupled haCC(6), $C$ dominates over $B$ at higher photo-electron energies. This can be attributed to interchannel coupling.

\subsection{Elliptically polarized ionizing field}

The upper panels of figure \ref{fig:800ellip} show total ionization spectra at 800 nm with elliptically polarized ionizing fields in XY and XZ planes and with 
the molecular axis along z axis. The pulses have been chosen to have peak fields in y,z directions respectively.
The spectra exhibit ATI peaks with an overall peak at about twice the pondermotive energy ($U_p \approx 0.22$ a.u) and with
no rescattering plateau. The channel analysis shows that the contribution of the excited state channels is about 10\% when the field is in XY plane and
about 5\% when the field is in XZ plane.  These relative contributions of the excited channels is in accordance with the findings in previous section with
linearly polarized fields. In the XY case, the relevant excited channel contribution comes from $A$ state while in the XZ case, $B$ channel dominates $A$ channel.
This stems from our choice of the peak field direction and the orbital symmetry. If the peak field was chosen to be along X (or Y) axis, $A$ channel would
dominate $B$ in both the cases.

\subsection{Angle resolved spectra}

\begin{figure*}[htbp]
\begin{minipage}{0.19\linewidth}
   \centering
   \includegraphics[width=0.95\linewidth]{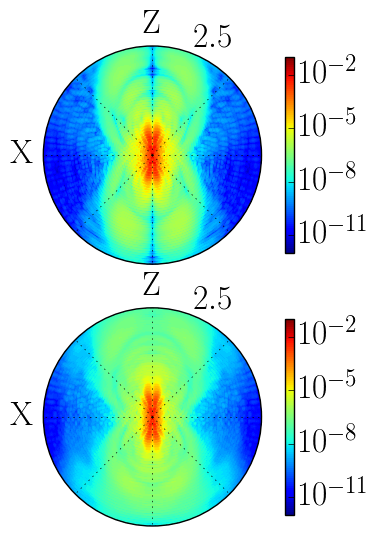}     
(a) $\alpha=0^\circ$
\end{minipage}
\begin{minipage}{0.19\linewidth}
   \centering
   \includegraphics[width=0.95\linewidth]{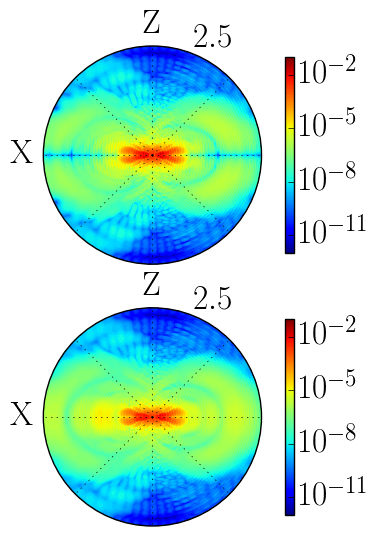}         
(b) $\alpha=90^\circ$
\end{minipage}
\begin{minipage}{0.19\linewidth}
   \centering
   \includegraphics[width=0.95\linewidth]{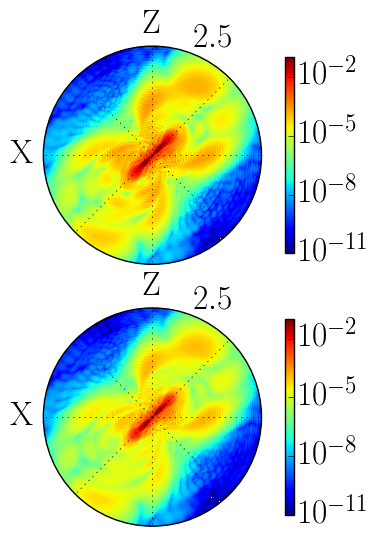}         
(c) $\alpha=45^\circ$
\end{minipage}
\begin{minipage}{0.19\linewidth}
   \centering
   \includegraphics[width=0.95\linewidth]{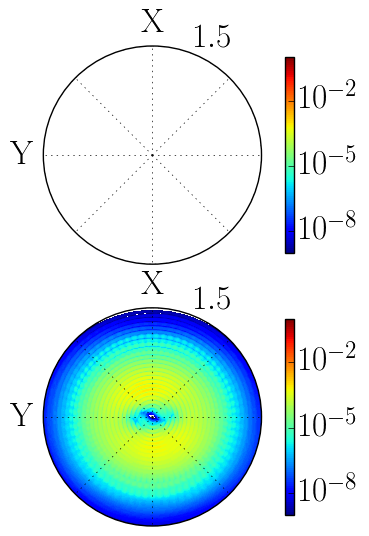}  
(d) $\epsilon=0.87$, XY
\end{minipage}
\begin{minipage}{0.19\linewidth}
   \centering
   \includegraphics[width=0.95\linewidth]{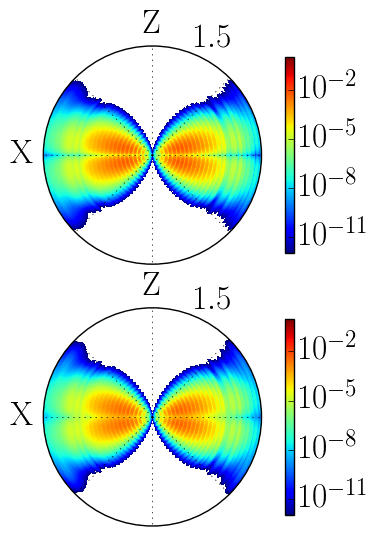} 
(e) $\epsilon=0.87$, XY
\end{minipage}
\begin{minipage}{0.19\linewidth}
   \centering
   \includegraphics[width=0.95\linewidth]{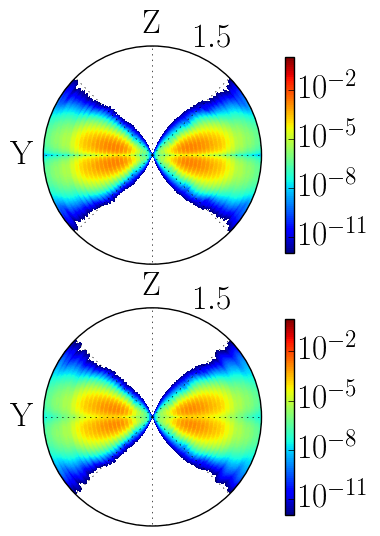}
(f) $\epsilon=0.87$, XY
\end{minipage}
\begin{minipage}{0.19\linewidth}
   \centering
   \includegraphics[width=0.95\linewidth]{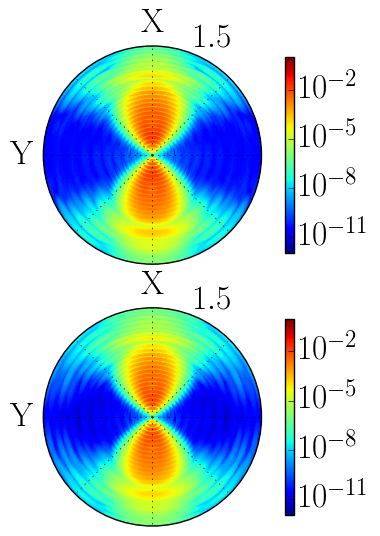}  
(g) $\epsilon=0.87$, XZ
\end{minipage}
\begin{minipage}{0.19\linewidth}
   \centering
   \includegraphics[width=0.95\linewidth]{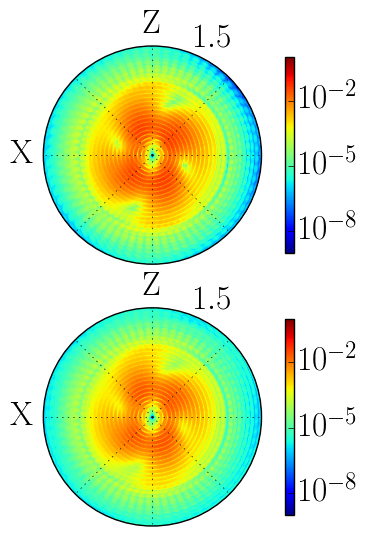} 
(h) $\epsilon=0.87$, XZ
\end{minipage}
\begin{minipage}{0.19\linewidth}
   \centering
   \includegraphics[width=0.95\linewidth]{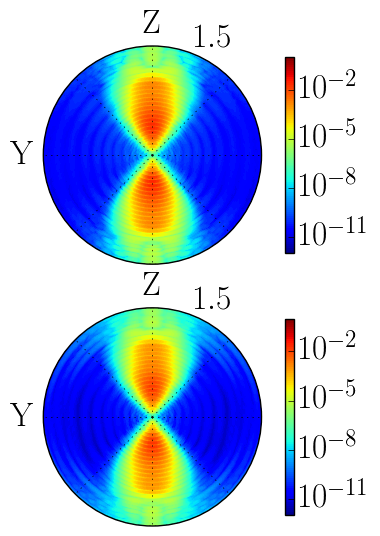}
(i) $\epsilon=0.87$, XZ
\end{minipage}
\caption{Cuts of angle, energy resolved spectra along various coordinate planes at 800 nm. The upper panel of each subplot corresponds to the ground state calculation (haCC(2)) and
the lower corresponds to the largest coupled channels calculation in each case. The molecular axis is fixed along Z axis.
Subplots (a)-(c) show spectra in the XZ plane with linearly polarized fields with polarization directions that subtend angles of $0^\circ$, $90^\circ$ and $45^\circ$ with
respect to Z axis. In the case of elliptic polarization, we show cuts along all the coordinate planes. Subplots (d)-(f) are the spectra when the elliptically
polarized field is in the XY plane and (g)-(i) are the spectra when the field is in the XZ plane. The upper panel of (d) is empty indicating zero emission and the white regions in
(e),(f) subplots result from masking of the spectral values $< 10^{-12}$.}
\label{fig:800angres}
\end{figure*}

In order to demonstrate the further potential of our method, we show in figure \ref{fig:800angres} angle resolved spectra summed over all the ionic channels
at 800 nm.
The spectra are only converged to a relative accuracy of 60\% with respect to the single electron basis ($|i\rangle$) parameters. 
We briefly mention below the qualitative features arising from symmetry and that are not subjected to the questions of quantitative convergence.

In the case of linear polarization, the cut along the XZ plane in presented. The spectra obtained from ground state channel calculation (haCC(2)) exhibit a nodal
structure for $\alpha=0^\circ, 90^\circ$ due to the
nodal planes in the highest occupied molecular orbital (HOMO). The contributions from the excited state channels lead to a wash out of the nodal structure. A similar
behavior is observed at 400 nm. 
When $\alpha=45^\circ$, selective enhancements are seen in the rescattering region, which we attribute to the structure of HOMO, which is the strongest ionization
channel at these parameters.

In the case of elliptic polarization, symmetry along molecular axis is not present and, we present cuts of the angle, energy resolved spectra in all the coordinate planes
XY, XZ and YZ. When the field is in the XZ plane, haCC(2) and haCC(5) results are qualitatively similar. The differential spectrum in XZ plane exhibit a four lobe structure
reflecting the shape of HOMO. The XY and YZ cuts exhibit a directional emission in X and Z directions respectively reflecting the laser polarization.
When the field is in XY plane, the spectra from haCC(2) and haCC(5) are qualitatively similar in XZ, YZ planes, again with a directional emission in accordance with
laser polarization. In the XY plane, however, the haCC(2) calculation shows zero emission resulting from the node in HOMO. When excited channels are included
contributions appear.

Obtaining better quantitative convergence and an analysis of the interference structures will be a topic of future study.

\subsection{Discussion}

Corroborating, it can be concluded that the nature of photoelectron spectra is determined by three factors: ionization thresholds, orbital symmetries
and interchannel coupling in the order of precedence. When total ionization yield to a specific channel is considered, ionization threshold is the determining
factor. In every example considered, the ground state channel has the highest yield. When differential spectra are considered, angle resolved or the rescattering
plateau that is sensitive to wavefunction in the plane of polarization, orbital symmetries come into the picture. That is, when an energetically favorable
channel has a node, the excited channel contribution becomes important. Finally, when a set of channels are similarly favored by the ionization thresholds and symmetry,
interchannel coupling plays a role. This is manifested in the case of $\alpha=0^\circ$ at 800 nm.
We do not observe the $X$, $B$ channel coupling proposed in the context of high harmonic spectra \cite{Smirnova2009}. A possible explanation would be that the 
photoelectron spectra and high harmonic spectra present different aspects of strong field dynamics.

Ionization in the case of linear polarization involves direct ionization along with contributions from recollision, where as ionization in the 
elliptic case is purely direct ionization. This is supported by the absence of the plateau feature that is a signature of recollision.
Comparing spectra from linear and elliptic case, it can be concluded that the excited channel contributions appear as a result of only the initial tunnel
ionization and recollision plays no role in populating these states. The relative dominance of the excited channels in the plateau region can be attributed to the
increase or decrease in the probability for recollision resulting from orbital symmetries.

The knowledge of orbital symmetries combined with molecular alignment technologies can be utilized to selectively populate a particular channel in the rescattering plateau. 
In the context of the LIED experiments, that aim to extract nuclear dynamics from the rescattering plateau, our results expose the
possibility to image channel specific dynamics. This is however subjected to corrections from nuclear motion.

\section{Conclusion}

We report first fully differential photoelectron spectra for a polyatomic molecule using a multi-electron description of the molecular
wavefunction. A coupled channels treatment is used to describe the molecular electronic structure and the time dependent surface 
flux method is used to compute the photoelectron spectra. The method is now accessible to studying strong field ionization of
linear polyatomics with arbitrary polarizations. Ionization of CO$_2$ by linearly and elliptically polarized laser fields with central wavelengths of
400 nm and 800 nm was considered.

Our calculations reveal that ionization to excited state channels is on the order of 5-10\%. While they may be neglected when considering
total yields, these states can dominate the ground state channel at certain momenta or orientations, where the ground state channel is prohibited for symmetry reasons.
This can lead to a host of interesting situations: (1) When considering angle resolved spectra, the nodal planes one expects from the
symmetry of the ground state channel do not appear. (2) In the momentum resolved but angle integrated spectra, specific excited state channels
can dominate in the rescattering plateau region depending on the molecular orientation. The rescattering process is sensitive to the electron
density in the polarization plane and a nodal plane in the energetically favorable channel leads to the relative dominance of an excited state channel.
When more than one channel become similarly favorable by energy and symmetry arguments, their sum can lead to a wash out of the ATI structure and also
allow the channels to dynamically couple during the ionization process.
In the context of the LIED experiments, that aim to extract dynamics from the rescattering plateau, our results with linear polarization at 800 nm
show that, one can potentially image channel specific nuclear dynamics by choosing a particular molecular orientation.

The qualitative nature of the differential spectra are defined by ionization thresholds, symmetries and channel coupling in the order of
precedence. These conclusions would also apply to other small and non polar molecules that exhibit bound state single ionic spectra similar to CO$_2$.

\section*{Acknowledgments}
The authors acknowledge financial support from the German excellence cluster: Munich Advanced Photonics and by the QUTIF (DFG Priority Program 1840).

\bibliography{molecularSpectra}{}
\bibliographystyle{apsrev4-1}

\end{document}